\documentclass[twocolumn,showpacs,preprintnumbers,a4paper,amsmath,amssymb,aps,prl,superscriptaddress]{revtex4-1}
\usepackage{epsfig}
\usepackage{graphicx}
\usepackage{dcolumn}
\usepackage{bm}
\usepackage{color} 

\begin{document}

\title{
Relation between $c$-$f$ hybridization and magnetic ordering in CeRu$_2$Al$_{10}$:\\
An optical conductivity study of Ce(Ru$_{1-x}$Rh$_x$)$_2$Al$_{10}$ ($x\leq0.05$)
}
\author{Shin-ichi Kimura}
\email{kimura@fbs.osaka-u.ac.jp}
\affiliation{FBS and Department of Physics, Osaka University, Suita, Osaka 565-0871, Japan}
\author{Hiroshi Tanida}
\affiliation{Department of Quantum Matter, ADSM, Hiroshima University, Higashi-Hiroshima, Hiroshima 739-8530, Japan}
\author{Masafumi Sera}
\affiliation{Department of Quantum Matter, ADSM, Hiroshima University, Higashi-Hiroshima, Hiroshima 739-8530, Japan}
\author{Yuji Muro}
\affiliation{Department of Liberal Arts and Sciences, Toyama Prefectural University, Toyama 939-0398, Japan}
\author{Toshiro Takabatake}
\affiliation{Department of Quantum Matter, ADSM, Hiroshima University, Higashi-Hiroshima, Hiroshima 739-8530, Japan}
\affiliation{Institute for Advanced Materials Research, Hiroshima University, Higashi-Hiroshima, Hiroshima 739-8530, Japan}
\author{Takashi Nishioka}
\affiliation{Graduate School of Integrates Arts and Science, Kochi University, Kochi 780-8520, Japan}
\author{Masahiro Matsumura}
\affiliation{Graduate School of Integrates Arts and Science, Kochi University, Kochi 780-8520, Japan}
\author{Riki Kobayashi}
\altaffiliation{Present address: Department of Physics and Earth Sciences, University of the Ryukyus, Okinawa 903-0213, Japan}
\affiliation{Graduate School of Integrates Arts and Science, Kochi University, Kochi 780-8520, Japan}
\date{\today}
\begin{abstract}
A Kondo semiconductor CeRu$_2$Al$_{10}$ with an orthorhombic crystal structure shows an unusual antiferromagnetic ordering at rather high temperature $T_0$ of 27.3~K, which is lower than the Kondo temperature $T_{\rm K}\sim$~60~K.
In optical conductivity [$\sigma(\omega)$] spectra that directly reflect electronic structure, the $c$-$f$ hybridization gap between the conduction and $4f$ states is observed at around 40~meV along the three principal axes.
However, an additional peak at around 20~meV appears only along the $b$ axis.
With increasing $x$ to 0.05 in Ce(Ru$_{1-x}$Rh$_x$)$_2$Al$_{10}$, the $T_0$ decreases slightly from 27.3~K to 24~K, but the direction of the magnetic moment changes from the $c$ axis to the $a$ axis.
Thereby, the $c$-$f$ hybridization gap in the $\sigma(\omega)$ spectra is strongly suppressed, but the intensity of the 20-meV peak remains as strong as for $x=0$.
These results suggest that the change of the magnetic moment direction originates from the decreasing of the $c$-$f$ hybridization intensity.
The magnetic ordering temperature $T_0$ is not directly related to the $c$-$f$ hybridization but is related to the charge excitation at 20 meV observed along the $b$ axis.
\end{abstract}

\pacs{71.27.+a, 78.20.-e}
%
%
%
\maketitle
%
%
Among rare-earth compounds, cerium (Ce) and ytterbium (Yb) compounds have various ground states such as nonmagnetic heavy fermion (HF) state due to the Kondo effect and magnetically ordered (MO) state.
The transition from HF to MO states is known to be determined as the energy valance between the Kondo effect and the Ruderman-Kittel-Kasuya-Yosida (RKKY) interaction.
The origin of the Kondo effect is the hybridization between conduction band and localized $4f$ states, namely $c$-$f$ hybridization, derived from the Anderson model~\cite{Hewson1993}, i.e., HF (MO) state appears in the region of the strong (weak) $c$-$f$ hybridization intensity.
Since the HF and MO phases are separated at a (quantum) critical point as described in the Doniach phase diagram, the HF and MO states do not appear coincidentally~\cite{Doniach1977}.
In the case of the strong $c$-$f$ hybridization intensity, some materials have a tiny energy gap, 
namely $c$-$f$ hybridization gap, 
at the Fermi level ($E_{\rm F}$), which are referred to Kondo insulators/semiconductors (KIs)~\cite{Aeppli1992,Takabatake1998}.

Owing to the strong $c$-$f$ hybridization intensity, KIs remain non-magnetic states at low temperatures.
However, recently found compounds Ce$M_2$Al$_{10}$ ($M$ = Ru, Os), anisotropic KIs with an orthorhombic structure~\cite{Thiede1998,Morrison2013}, show an antiferromagnetic ordering at the relatively higher temperature $T_0\sim$ 28~K~\cite{Strydom2009}, while they have an energy gap at $E_{\rm F}$ by the $c$-$f$ hybridization~\cite{Nishioka2009}.
Since this $T_0$ is beyond the de Genne scaling expected from the ordering temperatures of the other $RM_2$Al$_{10}$s ($R$ = rare-earth)~\cite{Muro2011,Adroja2013}, it is suggested that a novel interaction other than RKKY-type interaction should be adopted.

So far after the first report, so many experimental studies have been performed to reveal the origin of the anomalous magnetic ordering.
The results from these studies are summarized as follows:
There is an anisotropic $c$-$f$ hybridization in Ce$M_2$Al$_{10}$ ($M=$~Fe, Ru, Os).
The intensity of the $c$-$f$ hybridization in the $ac$ plane is stronger than that along the $b$ axis~\cite{Kimura2011-3,Sera2013,Tanida2014-3}.
Both the spin and charge gaps appear below the temperature slightly higher than $T_0$~\cite{Kimura2011-1}.
The magnetic moments direct to the $c$ axis at ambient pressure although the crystal field prefers the moment to orient to the $a$ axis~\cite{Strigari2012}.

In the $\sigma(\omega)$ spectra along the $a$ and $c$ axes, a $c$-$f$ hybridization gap ($\Delta_{c-f}$) is observed as same as these of conventional KIs.
On the other hand, along the $b$ axis, another gap ($\Delta_0$) of a charge excitation appears below $T_0$ in addition to $\Delta_{c-f}$~\cite{Kimura2011-1,Kimura2011-2}.
The additional energy gap possesses the similar structure of a superzone gap owing to the band folding.
The change of the electronic structure is considered to be related to the high $T_0$.

CeFe$_2$Al$_{10}$ has stronger $c$-$f$ hybridization intensity than Ce$M_2$Al$_{10}$ ($M=$~Ru, Os), and therefore has a non-magnetic ground state as in conventional KIs~\cite{Muro2009}.
Since CeFe$_2$Al$_{10}$ does not have $\Delta_0$ along the $b$ axis, the electronic structure of $\Delta_0$ along the $b$ axis appearing in other Ce$M_2$Al$_{10}$s seems to be strongly related to the magnetic ordering~\cite{Kimura2011-3}.
To clarify the relation of the magnetic ordering to the $c$-$f$ hybridization as well as to the charge excitation along the $b$ axis, the electronic structure in a system with weaker $c$-$f$ hybridization should be investigated.
By the substitution of Rh for Ru in CeRu$_2$Al$_{10}$, Ce(Ru$_{1-x}$Rh$_x$)$_2$Al$_{10}$, not only electrons are doped in the system with increasing $x$, but also the weaker $c$-$f$ hybridization can be realized~\cite{Tanida2014-2}.

With increasing $x$ in Ce(Ru$_{1-x}$Rh$_x$)$_2$Al$_{10}$, $T_0$ monotonically decreases from 27.3~K to 24~K for $x=$~0.05 and disappears at $x \sim 0.3$~\cite{Kobayashi2013}.
From magnetic susceptibility measurements, Kondo, Tanida and coworkers reported that 
the direction of the magnetic moment is strongly related to the $c$-$f$ hybridization.
It was proposed that the $c$-$f$ hybridization intensity along the $a$ axis drastically decreases by the Rh substitution at $x\sim0.05$~\cite{Kondo2013,Tanida2014-2}.
However, it is not clear why the $T_0$ remains up to $x \sim 0.3$ although the $c$-$f$ hybridization intensity is strongly suppressed already at $x\sim0.05$.

In this paper, we report the variation of the $\sigma(\omega)$ spectra as well as the electronic structure of Ce(Ru$_{1-x}$Rh$_x$)$_2$Al$_{10}$ ($x=0$, 0.03, 0.05).
The results demonstrate that $\Delta_{c-f}$ is strongly suppressed in between $x=0.03$ and $0.05$, suggesting that the $c$-$f$ hybridization intensity drastically decreases.
In contrast, the intensity of $\Delta_0$ hardly decreases and still remains up to $x=0.05$.
This change in $\Delta_0$ is in parallel with that of $T_0$.
These findings suggest that the magnetic ordering is not directly related to the $c$-$f$ hybridization but related to the charge excitations along the $b$ axis at around 20~meV.

%
\begin{figure}[t]
\begin{center}
\includegraphics[width=0.40\textwidth]{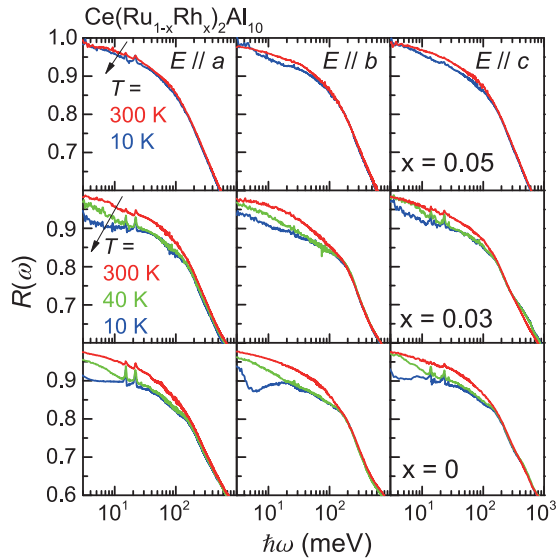}
\end{center}
\caption{
(Color online)
Temperature-dependent reflectivity [$R(\omega)$] spectra of Ce(Ru$_{1-x}$Rh$_x$)$_2$Al$_{10}$ ($x=0$, 0.03, 0.05) along all principal axes.
The $R(\omega)$ spectra of $x=0$ are referred from Ref.~\cite{Kimura2011-2}.
}
\label{fig:reflectivity}
\end{figure}
Single-crystalline samples of Ce(Ru$_{1-x}$Rh$_x$)$_2$Al$_{10}$ ($x=0.03, 0.05$) were synthesized by the Al-flux method~\cite{Muro2010-2} and the surfaces were well-polished using 0.3~$\mu$m grain-size Al$_{2}$O$_{3}$ lapping film sheets for the optical reflectivity [$R(\omega)$] measurements.
Near-normal incident polarized $R(\omega)$ spectra were acquired in a very wide photon-energy range of 2~meV -- 30~eV to ensure accurate Kramers-Kronig analysis (KKA)~\cite{Kimura2013}.
Martin-Puplett and Michelson type rapid-scan Fourier spectrometers (JASCO Co. Ltd., FARIS-1 and FTIR6100) were used at the photon energy $\hbar\omega$ regions of 2~--~30~meV and 5~meV~--~1.5~eV, respectively, with a feed-back positioning system to maintain the overall uncertainty level less than $\pm$0.5~\% in the temperature range of 10~--~300~K~\cite{Kimura2008}.
To obtain the absolute $R(\omega)$ values, the samples were evaporated {\it in situ} with gold, whose spectrum was then measured as a reference.
The obtained polarized $R(\omega)$ spectra of all Ce(Ru$_{1-x}$Rh$_x$)$_2$Al$_{10}$s in the photon energy range of 2~meV -- 1~eV are plotted in Fig.~\ref{fig:reflectivity} with those of CeRu$_2$Al$_{10}$ ($x=0$) referred from Ref.~\cite{Kimura2011-2}.
At $T=300$~K, $R(\omega)$ of CeRu$_2$Al$_{10}$ was only measured for energies 1.2--30~eV by using synchrotron radiation, and connected to the spectra of all Ce(Ru$_{1-x}$Rh$_x$)$_2$Al$_{10}$s for the KKA.
In order to obtain $\sigma(\omega)$ via KKA of $R(\omega)$, the spectra were extrapolated below 2~meV with a Hagen-Rubens function, and above 30~eV with a free-electron approximation $R(\omega) \propto \omega^{-4}$~\cite{DG}.


\begin{figure}[t]
\begin{center}
\includegraphics[width=0.4\textwidth]{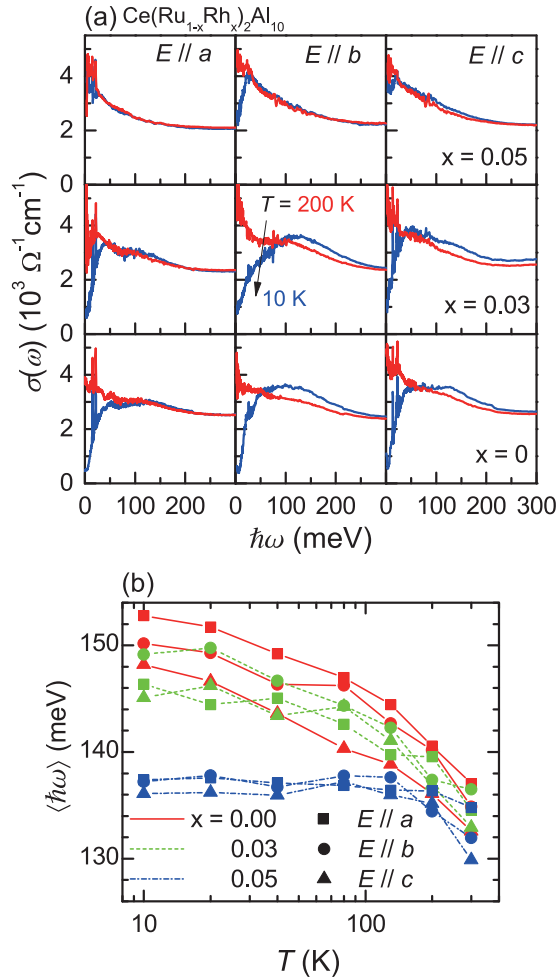}
\end{center}
\caption{
(Color online)
(a) Temperature-dependent optical conductivity [$\sigma(\omega)$] spectra of Ce(Ru$_{1-x}$Rh$_x$)$_2$Al$_{10}$ ($x=0$, 0.03, 0.05) along all principal axes in the photon energy region below 300~meV.
(b) The centers of gravity [$\langle\hbar\omega\rangle$] of the $\sigma(\omega)$ spectra are plotted as a function of temperature.
}
\label{fig:OCwide}
\end{figure}
$\sigma(\omega)$ spectra for $\hbar\omega\leq$~300~meV are shown in Fig.~\ref{fig:OCwide}(a).
The wide-range $R(\omega)$ spectrum in Fig.~\ref{fig:reflectivity} as well as the $\sigma(\omega)$ spectrum hardly changes in the range of $\hbar\omega>$~300~meV by temperature.
Therefore the $\sigma(\omega)$ spectrum has temperature dependence only in the range of $\hbar\omega\leq$~300~meV.
A broad peak observed at about 100~meV in $x=0$ originates from the optical transition from the valence band to the unoccupied Ce~$4f_{5/2}$ state~\cite{Kimura2009}.
Both the peak intensity and the temperature dependence, which denote to the $c$-$f$ hybridization character, imply that the $c$-$f$ hybridization intensity decreases with increasing $x$.
To clarify the temperature dependence, the center of gravity 
\[
\langle\hbar\omega\rangle=\int_{0}^{\omega_{max}} \hbar\omega \sigma(\omega) d\omega/ \int_{0}^{\omega_{max}} \sigma(\omega) d\omega
\]
below 300~meV ($=\hbar\omega_{max}$) is plotted in Fig.~\ref{fig:OCwide}(b).
The figure shows that the $\langle\hbar\omega\rangle$s of $x=0$ and $0.03$ along all principle axes have the same temperature dependence as one another, however the $\langle\hbar\omega\rangle$ of $x=0.05$ is almost constant with temperature.
These results suggest that the $c$-$f$ hybridization intensity of $x=0.03$ is as same as that of $x=0$, but that of $x=0.05$ is strongly reduced from that of $x=0.03$.
The strong suppression of the $c$-$f$ hybridization at the low level substitution at $x=0.05$ implies that this material could be a Kondo lattice system, which is different from a single-site Kondo system like YbB$_{12}$~\cite{Okamura2000}.
In between $x=0.03$ and $0.05$, $T_0$ is almost constant but the direction of the magnetic moment changes from the $c$ axis to $a$ axis.
Since the magnetic moment along the $a$ axis is larger than that along the $c$ axis, the flop of the magnetic moment direction is considered to originate from the decreasing of the $c$-$f$ hybridization intensity.
Namely, the magnetic moment direction along the $c$ axis of CeRu$_2$Al$_{10}$ at ambient pressure is owing to the strong $c$-$f$ hybridization intensity.
However, since the $T_0$ is almost constant for $x\leq0.05$, the magnetic moment direction seems to be not important to the magnetic ordering, but the direction is determined by the anisotropic $c$-$f$ hybridization intensity.

\begin{figure}[b]
\begin{center}
\includegraphics[width=0.4\textwidth]{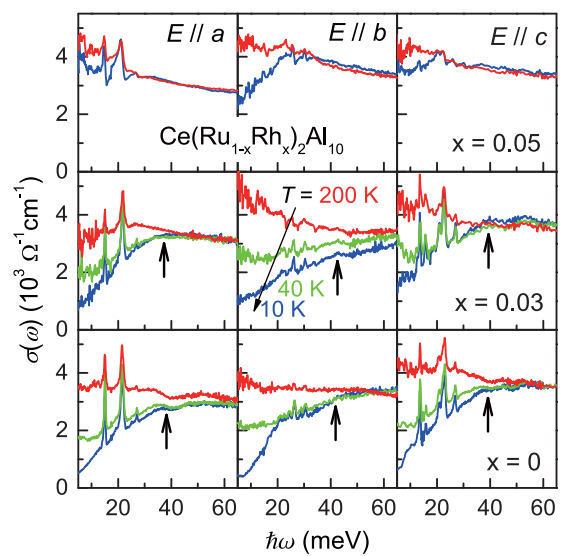}
\end{center}
\caption{
(Color online)
Temperature-dependent optical conductivity [$\sigma(\omega)$] spectra of Ce(Ru$_{1-x}$Rh$_x$)$_2$Al$_{10}$ ($x=0$, 0.03, 0.05) along all principal axes in the photon energy range below 65~meV.
Vertical arrows indicate the optical transitions across the $c$-$f$ hybridization gap.
}
\label{fig:OCnarrow}
\end{figure}
Next, we focus on the energy gap structures.
The temperature dependence of the $\sigma(\omega)$ spectrum for $\hbar\omega\leq$~65~meV is shown in Fig.~\ref{fig:OCnarrow}.
In this region, sharp peaks due to transverse optical phonons are observed at around 20~meV.
These peaks become broad with increasing $x$, indicating that the life time of the phonons become shorter with decreasing $x$.
Note that the peak broadening does not originate from the disorder of substituted elements because the electronic specific heat of CeRu$_2$Al$_{10}$ is not varied by the substitution of elements with the same valence~\cite{Tanida2014-2}.
This originates in the increase of the interaction between carriers and phonons.
It means that the carrier number reasonably increases with increasing $x$.
Since the shoulder structures pointed by thick arrows at lower temperatures originate from the optical transition across the $c$-$f$ hybridization gap, 
the energy of the shoulder edge corresponds to the twice of the $c$-$f$ hybridization energy~\cite{Iizuka2010}.
The shoulder structure does not change from $x=0$ to $0.03$.
However, it disappears at $x=0.05$, indicating that the $c$-$f$ hybridization is strongly suppressed at $x=0.05$.
This result is consistent with the temperature-independent $\langle\hbar\omega\rangle$ of the mid-IR peak in Fig.~\ref{fig:OCwide} and also the result of the high-field magnetic susceptibility~\cite{Kondo2013}.
At $x=0.05$, the energy gap along the $a$ and $c$ axes obviously disappears and then the $\sigma(\omega)$ spectra change to metallic.
On the other hand, along the $b$ axis, a broad peak structure at $\hbar\omega\sim$~20~meV remains.
The peak energy is obviously different from the $c$-$f$ hybridization gap size of about 40~meV but is as same as the 20-meV peak of $x=0$ and $0.03$, i.e., the $c$-$f$ hybridization gap disappears but the 20-meV peak remains for $x=0.05$.
The appearance of the peak at $\hbar\omega\sim$~20~meV for $x=0$ has been discussed to be related to the magnetic ordering at $T_0$~\cite{Kimura2011-1,Kimura2011-2,Kimura2012}.
Since the magnetic ordering occurs at the similar temperature of $T_0$, the 20-meV peak is confirmed to be related to the magnetic ordering.
Note that a peak at $\hbar\omega\sim$~20~meV is recognized for the $c$ axis of $x=0.05$.
The peak can be regarded as the absorption peak due to TO phonons because of the same energy of the peaks in $x=0$ and $0.03$.

\begin{figure}[t]
\begin{center}
\includegraphics[width=0.4\textwidth]{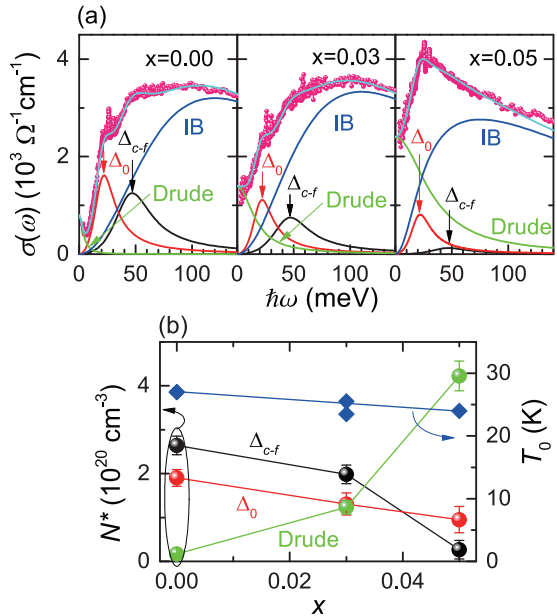}
\end{center}
\caption{
(Color online)
(a) $\sigma(\omega)$ spectra along the $b$ axis at 10~K and the fitted lines of one Drude and three Lorentz functions.
The three Lorentz functions are assumed as a peak appearing below $\sim T_0$, the interband transition (IB) in the $c$-$f$ hybridization gap, and higher interband transition from lower energy side.
(b) Obtained effective electron numbers of the Drude, $\Delta_0$, and $\Delta_{c-f}$ components and the magnetic ordering temperature $T_0$ as a function of the Rh-substitution $x$.
Note that, at $x=0.03$, two magnetic ordering temperature at 23.5 and 25.5~K reported in Ref.~\cite{Tanida2014-2} are plotted.
}
\label{fig:baxis}
\end{figure}
The $x$-dependence of the electronic structure along the $b$ axis is discussed because the $b$-axis electronic structure plays an important role in the magnetic ordering and the $\sigma(\omega)$ spectra as well as the electronic structures along the $a$ and $c$ axes do not change across $T_0$~\cite{Kimura2011-1}
.
In Figure~\ref{fig:baxis}(a), the $\sigma(\omega)$ spectra of $x=0$, 0.03, 0.05 along the $b$ axis at $T=$~10~K are fitted by using the combination of four Lorentz functions:
\[
\sigma(\omega)=\sum_{i=0}^3\frac{N^*_ie^2\tau_i\omega^2}{m_0{(\omega^2_i-\omega^2)^2\tau^2_i+\omega^2}}.
\]
Here, $i=0$ indicates a Drude component with $\omega_0=0$ presenting the existence of carriers.
$m_0$ denotes the electron rest mass, $N^*_0, N^*_1, N^*_2$, and $N^*_3$ are the effective electron numbers and $\tau_0, \tau_1, \tau_2$, and $\tau_3$ the relaxation times of Drude, $\Delta_0$, $\Delta_{c-f}$, and higher interband transition, respectively.
$\omega_1$, $\omega_2$, and $\omega_3$ are the resonant frequencies of the Lorentz absorption.
Fitting results along the $a$ and $c$ axes are shown in the supplemental Figs.~S1 and S2.
The fitting was performed by using fixed parameters of the peak centers of $\omega_1$ and $\omega_2$ and the widths of $\tau_1$ and $\tau_2$ for the $\Delta_0$ and $\Delta_{c-f}$, respectively, in the assumption that the gap energies and shapes are reasonably constant with different $x$.
The higher interband transitions were fitted at the slope at $\hbar\omega\sim$~140~meV.
The obtained values of the effective electron number $N^*$ and $T_0$~\cite{Tanida2014-2} as a function of $x$ are plotted in Fig.~\ref{fig:baxis}(b).
Note that, at $x=0.03$, two magnetic ordering temperatures at 23.5 and 25.5~K that have been reported~\cite{Tanida2014-2} are plotted in Fig.~\ref{fig:baxis}(b).

From Figure~\ref{fig:baxis}, the following results can be obtained:
With increasing $x$, the intensity of the Drude component increases, the carrier density increases significantly between $x=0.03$ and $0.05$.
In this $x$ region, on the other hand, the $\Delta_{c-f}$ peak intensity suddenly decreases.
These results suggest that the $c$-$f$ hybridization gap is suddenly collapsed and then the carrier density rapidly increases from $x=0.03$ to $0.05$.
This implies that the electronic structure does not rigidly shift by the Rh substitution.
On the other hand, the intensity of $\Delta_0$ monotonically and slightly decreases with increasing $x$.
The decrease of $\Delta_0$ is in parallel with the decreasing $T_0$.
Therefore, as discussed so far, not the $c$-$f$ hybridization but the electronic structure producing the $\Delta_0$ peak is strongly related to the magnetic ordering at $T_0$.

Finally, let us discuss the origin of the magnetic ordering.
We got the conclusion that the magnetic ordering is not directly related to the $c$-$f$ hybridization.
In the $\sigma(\omega)$ spectra, there is a charge excitation gap along the $b$ axis, in addition to the $c$-$f$ hybridization gap along all three axes.
The gap along the $b$ axis is similar to the gap structure with a peak observed in materials with charge-density waves and spin-density waves~\cite{Li2007,Dressel1997}, namely a superzone gap owing to a band folding.
Similar peak structure also appears in the $\sigma(\omega)$ spectrum of URu$_2$Si$_2$ at the hidden-order (HO) phase~\cite{Nagel2012}.
At the HO phase, the order parameter had not been revealed for a long time~\cite{Mydosh2011}.
However, recently the lattice symmetry breaking has been observed by using a high-resolution X-ray diffraction using synchrotron radiation~\cite{Tonegawa2014}.
Then, the peak structure in the $\sigma(\omega)$ spectrum in the HO phase is considered to be associated with a lattice distortion.
In CeOs$_2$Al$_{10}$, the superlattice structure has been observed by using an electron diffraction~\cite{Muro2010-2}.
However no other obvious result has been obtained yet.
Therefore the change of the $b$-axis electronic structure might originate from the same as that of the HO state of URu$_2$Si$_2$.
Future high-resolution X-ray diffraction and other precise experiments may reveal the lattice distortion along the $b$ axis of Ce$M_2$Al$_{10}$ systems, which is similar to that of the HO phase of URu$_2$Si$_2$.

%
To summarize, the relation between the $c$-$f$ hybridization intensity and the unusual magnetic ordering of CeRu$_2$Al$_{10}$ has been investigated by using the optical conductivity spectra of the electron doped Ce(Ru$_{1-x}$Rh$_x$)$_2$Al$_{10}$ ($x=$0, 0.03, 0.05).
Both the $c$-$f$ hybridization gap at $\hbar\omega\sim$~40~meV and the mid-IR peaks at around 100~meV were clearly observed in the range $x\leq$~0.03, but almost disappear at $x=0.05$.
This suggests that the $c$-$f$ hybridization is rapidly suppressed with increasing $x$ from $0.03$ to $0.05$.
Since the magnetic moment direction changes from the $c$ axis to $a$ axis at the Rh substitution content, the change is considered to originate from the decreasing of the $c$-$f$ hybridization intensity.
On the other hand, the magnetic ordering temperature $T_0$ slightly decreases from 27.3~K to 24~K.
The $x$ dependence is similar to that of the additional peak along the $b$ axis observed at around 20~meV.
Therefore, we conclude that the magnetic ordering is not directly related to the $c$-$f$ hybridization but is induced by the electronic structure producing the 20~meV-peak.

%
We would like to thank UVSOR staff members for their technical support.
Part of this work was supported by the Use-of-UVSOR Facility Program (BL6B, 2011-2012) of the Institute for Molecular Science.
This work was partly supported by JSPS KAKENHI Grant Numbers 22340107 and 26400363.


%


\setcounter{table}{0}
\setcounter{figure}{0}
\onecolumngrid

\newpage

\begin{center}
{\bf \Large
Supplementary Material for: Relation between $c$-$f$ hybridization and magnetic ordering in CeRu$_2$Al$_{10}$: 
An optical conductivity study of Ce(Ru$_{1-x}$Rh$_x$)$_2$Al$_{10}$ ($x\leq0.05$)
}
\end{center}

\begin{center}
Shin-ichi Kimura$^{1,*}$\email{kimura@fbs.osaka-u.ac.jp}, Hiroshi Tanida$^{2}$, Masafumi Sera$^{2}$, Yuji Muro$^{3}$,
Toshiro Takabatake$^{2,4}$, Takashi Nishioka$^{5}$, Masahiro Matsumura$^{5}$, Riki Kobayashi$^{5,+}$\\
$^1${\it FBS and Department of Physics, Osaka University, Suita, Osaka 565-0871, Japan}\\
$^2${\it Department of Quantum Matter, ADSM, Hiroshima University, Higashi-Hiroshima, Hiroshima 739-8530, Japan}\\
$^3${\it Department of Liberal Arts and Sciences, Toyama Prefectural University, Toyama 939-0398, Japan}\\
$^4${\it Institute for Advanced Materials Research, Hiroshima University, Higashi-Hiroshima, Hiroshima 739-8530, Japan}\\
$^5${\it Graduate School of Integrates Arts and Science, Kochi University, Kochi 780-8520, Japan}\\

\date{\today}
\end{center}

\twocolumngrid

\renewcommand{\thefigure}{S\arabic{figure}}
\renewcommand{\thetable}{S\arabic{table}}

\section{Fitting results along $a$ and $c$ axes}
Figures~S1 and S2 show Drude and Lorentz fitting results of Ce(Ru$_x$Rh$_{1-x}$)$_2$Al$_{10}$ ($x = 0, 0.03, 0.05$) at 10~K along the $a$ and $c$ axes, respectively.
In both axes, one Drude and two Lorentz components [$\Delta_{c-f}$ and a higher interband transition (IB)] are assumed because no peak feature appears at the energy of $\Delta_0$ along the $b$ axis as shown in Fig.~4.
Note that there is a peak at $\hbar\omega\sim$~20~meV for the $c$ axis of $x=0.05$.
The peak is not the same as $\Delta_0$ along the $b$ axis but originates from the TO phonons because of the same energy of the TO phonon peaks for $x=0$ and $0.03$ as shown in Fig.~3.
The intensity of the Drude component increases slightly from $x = 0$ to $0.03$ and rapidly from $0.03$ to $0.05$, and the $\Delta_{c-f}$ peak intensities along the both axes behave oppositely, which is the same way as along the $b$ axis.
These $x$ dependences are different from that of $T_0$~\cite{Tanida2014-2}.
Therefore the $\sigma(\omega)$ spectra as well as the electronic structures along the $a$ and $c$ axes are concluded not to be related to the magnetic ordering at $T_0$, which is the same as that in Ref.~\cite{Kimura2011-1}.

\begin{figure}
\begin{center}
\includegraphics[width=0.4\textwidth]{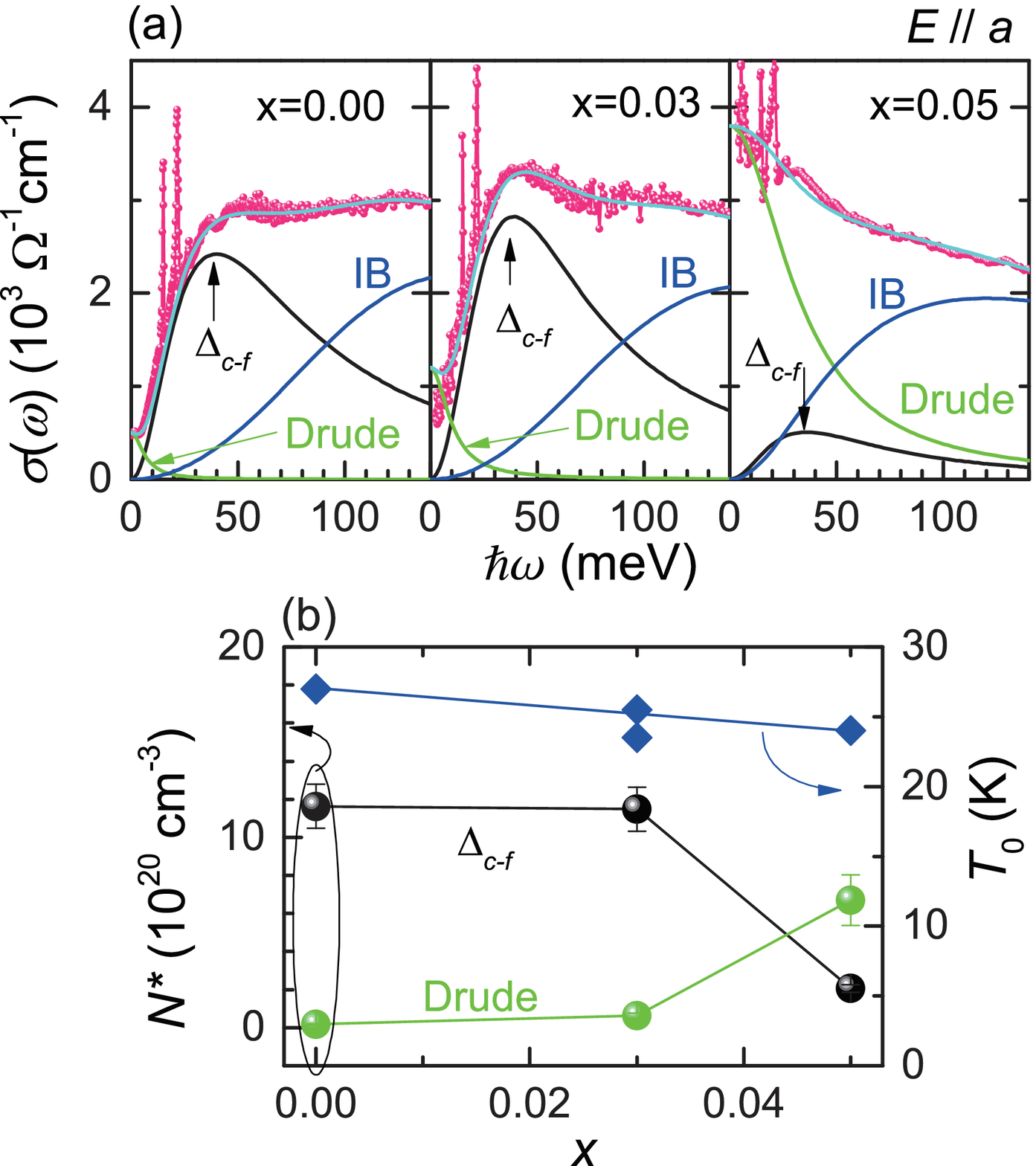}
\end{center}
\caption{
(Color online)
(a) $\sigma(\omega)$ spectra along the $a$ axis at 10~K and the fitted lines of one Drude and three Lorentz functions.
The two Lorentz functions are assumed as the interband transition in the $c$-$f$ hybridization gap ($\Delta_{c-f}$) and higher interband transition (IB) from lower energy side.
(b) Obtained effective electron numbers of the Drude and $\Delta_{c-f}$ components and the magnetic ordering temperature $T_0$ as a function of the Rh-substitution $x$.
Note that, at $x=0.03$, two magnetic ordering temperature at 23.5 and 25.5~K reported in Ref.~\cite{Tanida2014-2} are plotted.
}
\label{fig:aaxis}
\end{figure}
\begin{figure}
\begin{center}
\includegraphics[width=0.4\textwidth]{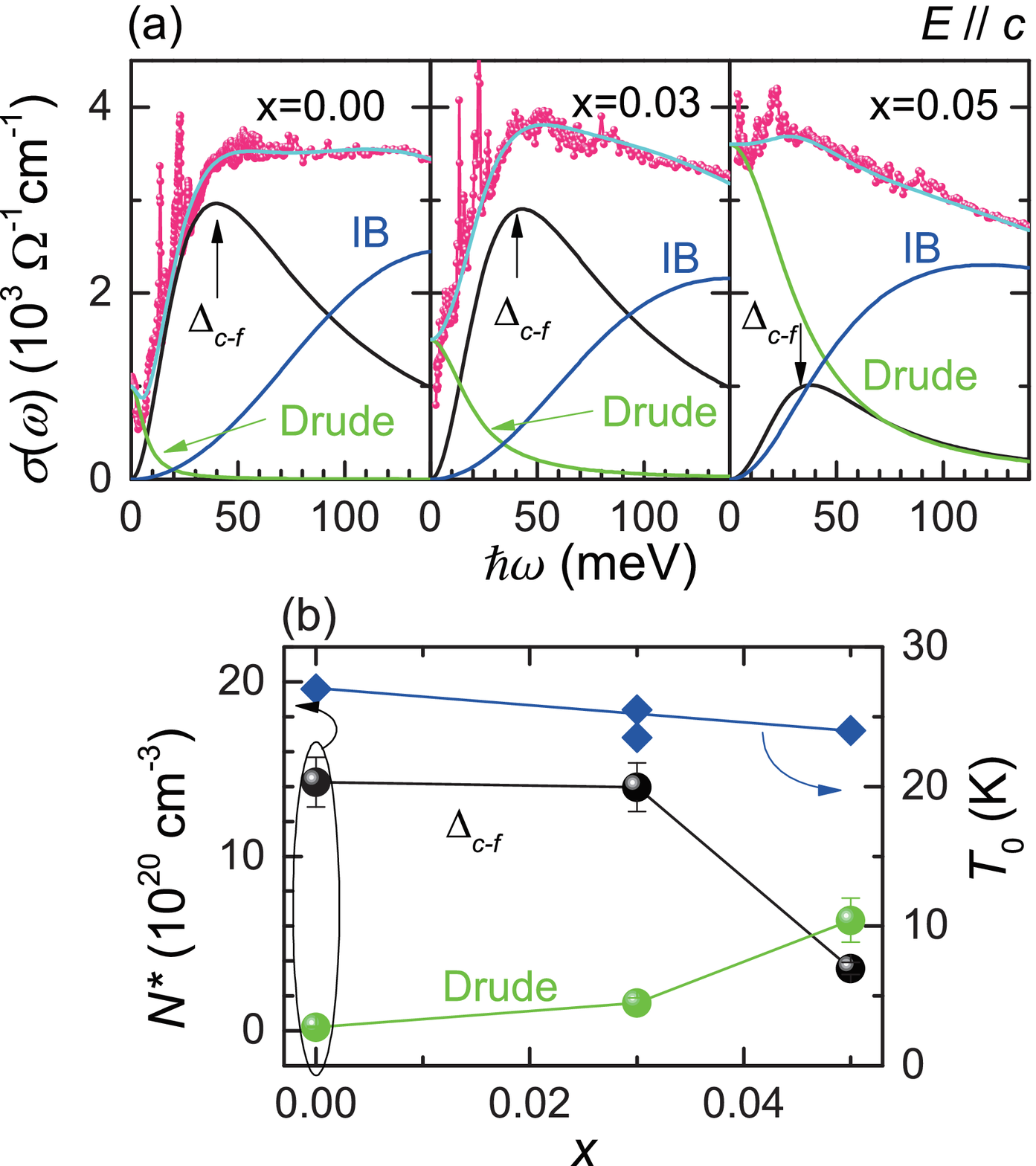}
\end{center}
\caption{
(Color online)
Same as Fig.~\ref{fig:aaxis}, but for $E \parallel c$.
}
\label{fig:baxis}
\end{figure}


%


\begin{thebibliography}{99}
%
\bibitem{Hewson1993}
A. C. Hewson: {\it The Kondo Problem to Heavy Fermions} (Cambridge University Press, Cambridge, 1993).
%
\bibitem{Doniach1977}
S. Doniach, Physica {\bf 91}, 231 (1977). 
%
\bibitem{Aeppli1992}
G. Aeppli and Z. Fisk, Comments Condens. Matter Phys. {\bf 16}, No. 3, 155 (1992), and references therein.
%
\bibitem{Takabatake1998}
T. Takabatake, F. Iga, T. Yoshino, Y. Echizen, K. Katoh, K. Kobayashi, M. Higa, N. Shimizu, Y. Bando, G. Nakamoto, H. Fujii, K. Izawa, T. Suzuki, T. Fujita, M. Sera, M. Hiroi, K. Maezawa, S. Mock, H. v. L\"ohneysen, A. Br\"uckl, K. Neumaier, and K. Andres, J. Magn. Magn. Mater. {\bf 177-181}, 277 (1998), and references therein.
%
\bibitem{Thiede1998}
V. M. T. Thiede, T. Ebel, and W. Jeitschko, J. Mater. Chem. {\bf 8}, 125 (1998).
%
\bibitem{Morrison2013}
G. Morrison, N. Haldolaarachchige, C.-W. Chen, D. P. Young, E. Morosan, and J. Y. Chan, Inorg. Chem. {\bf 52}, 3198 (2013).
%
\bibitem{Strydom2009}
A. M. Strydom, Physica B {\bf 404}, 2981 (2009).
%
\bibitem{Nishioka2009}
T. Nishioka, Y. Kawamura, T. Takesaka, R. Kobayashi, H. Kato, M. Matsumura, K. Kodama, K. Matsubayashi, and Y. Uwatoko, J. Phys. Soc. Jpn. {\bf 78}, 123705 (2009).
%
\bibitem{Muro2011}
Y. Muro, J. Kajino, T. Onimaru, and T. Takabatake, J. Phys. Soc. Jpn. {\bf 80}, SA021 (2011).
%
\bibitem{Adroja2013}
D. T. Adroja, A. D. Hillier, Y. Muro, T. Takabatake, A. M. Strydom, A. Bhattacharyya, A. Daoud-Aladin, and J. W. Taylor, Phys. Scr. {\bf 88}, 068505 (2013).
%
\bibitem{Kimura2011-3}
S. Kimura, Y. Muro, and T. Takabatake, J. Phys. Soc. Jpn. {\bf 80}, 033702 (2011).
%
\bibitem{Sera2013}
M. Sera, D. Tanaka, H. Tanida, C. Moriyoshi, M. Ogawa, Y. Kuroiwa, T. Nishioka, M. Matsumura, J. Kim, N. Tsuji, and M. Takata, J. Phys. Soc. Jpn. {\bf 82}, 024603 (2013).
%
\bibitem{Tanida2014-3}
H. Tanida, H. Nohara, M. Nakamura, M. Sera, T. Terashima, S. Uji, T. Nishioka, and M. Matsumura, JPS Conf. Proc. {\bf 3}, 011073 (2014).
%
\bibitem{Kimura2011-1}
S. Kimura, T. Iizuka, H. Miyazaki, A. Irizawa, Y. Muro, and T. Takabatake, Phys. Rev. Lett. {\bf 106}, 056404 (2011).
%
\bibitem{Strigari2012}
F. Strigari, T. Willers, Y. Muro, K. Yutani, T. Takabatake, Z. Hu, Y.-Y. Chin, S. Agrestini, H.-J. Lin, C. T. Chen, A. Tanaka, M. W. Haverkort, L. H. Tjeng, and A. Severing, Phys. Rev. B {\bf 86}, 081105(R) (2012).
%
\bibitem{Kimura2011-2}
S. Kimura, T. Iizuka, H. Miyazaki, T. Hajiri, M. Matsunami, T. Mori, A. Irizawa, Y. Muro, J. Kajino, and T. Takabatake, Phys. Rev. B {\bf 84}, 165125 (2011).
%
\bibitem{Muro2009}
Y. Muro, K. Motoya, Y. Saiga, and T. Takabatake, J. Phys. Soc. Jpn. {\bf 78}, 083707 (2009).
%
\bibitem{Tanida2014-2}
H. Tanida, H. Nohara, M. Sera, T. Nishioka, M. Matsumura, and R. Kobayashi, Phys. Rev. B {\bf 90}, 165124 (2014).
%
\bibitem{Kobayashi2013}
R. Kobayashi, Y. Ogane, D. Hirai, T. Nishioka, M. Matsumura, Y. Kawamura, K. Matsubayashi, Y. Uwatoko, H. Tanida, and M. Sera, J. Phys. Soc. Jpn. {\bf 82}, 093702 (2013).
%
\bibitem{Kondo2013}
A. Kondo, K. Kindo, K. Kunimori, H. Nohara, H. Tanida, M. Sera, R. Kobayashi, T. Nishioka, and M. Matsumura, J. Phys. Soc. Jpn. {\bf 82}, 054709 (2013).
%
\bibitem{Muro2010-2}
Y. Muro, J. Kajino, K. Umeo, K. Nishimoto, R. Tamura, and T. Takabatake, Phys. Rev. B {\bf 81}, 214401 (2010).
%
\bibitem{Kimura2013}
S. Kimura and H. Okamura, J. Phys. Soc. Jpn. {\bf 82}, 021004 (2013).
%
\bibitem{Kimura2008}
S. Kimura, JASCO Report {\bf 50}, 6 (2008). [in Japanese]
%
\bibitem{DG}
M. Dressel and G. Gr\"uner, {\it Electrodynamics of Solids} (Cambridge University Press, Cambridge, UK, 2002).
%
\bibitem{Kimura2009}
S. Kimura, T. Iizuka, and Y. S. Kwon, J. Phys. Soc. Jpn. {\bf 78}, 013710 (2009).
%
\bibitem{Okamura2000}
H. Okamura, M. Matsunami, T. Inaoka, T. Nanba, S. Kimura, F. Iga, S. Hiura, J. Klijn, and T. Takabatake, Phys. Rev. B {\bf 62}, R13265 (2000).
%
\bibitem{Iizuka2010}
T. Iizuka, S. Kimura, A. Herzog, J. Sichelschmidt, C. Krellner, C. Geibel, and F. Steglich, J. Phys. Soc. Jpn. {\bf 79}, 123703 (2010).
%
\bibitem{Kimura2012}
S. Kimura, T. Iizuka, Y. Muro, J. Kajino, and T. Takabatake, J. Phys.: Conf. Ser. {\bf 391}, 012030 (2012). 
%
\bibitem{Li2007}
G. Li, W. Z. Hu, D. Qian, D. Hsieh, M. Z. Hasan, E. Morosan, R. J. Cava, and N. L. Wang, Phys. Rev. Lett. {\bf 99}, 027404 (2007).
%
\bibitem{Dressel1997}
M. Dressel, L. Degiorgi, J. Brickmann, A. Schwartz, and G. Gr\"uner, Physica B {\bf 230-232}, 1008 (1997).
%
\bibitem{Nagel2012}
U. Nagel, T. Uleksin, T. R\~o\~om, R. P. S. M. Lobo, P. Lejay, C. C. Homes, J. S. Hall, A. W. Kinross, S. K. Purdy, T. Munsie, T. J. Williams, G. M. Luke, and T. Timusk, Proc. Natl. Acad. Sci. USA {\bf 109}, 19161 (2012).
%
\bibitem{Mydosh2011}
J. A. Mydosh and P. M. Oppeneer, Rev. Mod. Phys. {\bf 83}, 1301 (2011).
%
\bibitem{Tonegawa2014}
S. Tonegawa, S. Kasahara, T. Fukuda, K. Sugimoto, N. Yasuda, Y. Tsuruhara, D. Watanabe, Y. Mizukami, Y. Haga, T. D. Matsuda, E. Yamamoto, Y. \^Onuki, H. Ikeda, Y. Matsuda, and T. Shibauchi, Nature Commun. {\bf 5}, 4188 (2014).
%
\end{thebibliography}

\begin{thebibliography}{99}
%
\bibitem{Tanida2014-2}
H. Tanida, H. Nohara, M. Sera, T. Nishioka, M. Matsumura, and R. Kobayashi, Phys. Rev. B {\bf 90}, 165124 (2014).
%
\bibitem{Kimura2011-1}
S. Kimura, T. Iizuka, H. Miyazaki, A. Irizawa, Y. Muro, and T. Takabatake, Phys. Rev. Lett. {\bf 106}, 056404 (2011).
%
\end{thebibliography}
\end{document}